# Inelastic scattering of neutrons by superconducting rings

A I Agafonov

Russian Research Center "Kurchatov Institute", Moscow, 123182 Russia

E-mail: aai@isssph.kiae.ru

**Abstract.** The differential cross section for the inelastic magnetic scattering of neutrons by superconducting rings is derived theoretically taking account of the interaction of the neutron magnetic moment with magnetic field created by the superconducting current. In this scattering process the neutron kinetic energy increases discretely and, respectively, the number of the magnetic flux quanta trapped in the ring, reduces. Quantitative calculations of the scattering cross section for cold neutrons is carried out for thin film rings from type-II superconductors with the film thickness $d < \lambda$ ($\lambda$ is the penetration depth).

**1. Introduction**

The magnetic interaction of neutrons with a scatterer is understood the interaction of the magnetic field generated by neutron, with currents of electronic transitions in target atoms. This interaction can also be presented in the completely identical form of the interaction of the neutron magnetic moment with the magnetic field created by the electronic currents of the scatterer atoms [1]. Various channels of the magnetic scattering of neutrons by solid and liquid targets were studied [2-4]. However, the interaction of neutrons with magnetic fields of superconducting rings, to our knowledge, has not previously been investigated.

Due to this inelastic scattering process, the macroscopic quantum state of the superconducting condensate in the ring can be changed by single neutron. To explain this process, consider a superconducting ring with a current $J_m = \dfrac{\Phi_m}{L}$, where $\Phi_m$ is the magnetic induction flux trapped in the ring, $m = \Phi_m / \Phi_0$ is the number of the superconducting flux quanta (fluxoids), $\Phi_0 = \dfrac{h}{2e}$ is the fluxoid and $L$ is the ring self-inductance. The superconducting current defined by the initial condensate wave function $\psi_m$ in the ring, creates the magnetic field. The neutron magnetic moment interacts with this field. Because the total energy of the superconducting ring $E_m \propto m^2$ in the state $\psi_m$, the kinetic energy of neutron can change only discretely depending on the final number $m_1$ of the magnetic flux quanta trapped in the ring. This inelastic scattering process must be accompanied by a transition of the superconducting condensate from the initial state $\psi_m$ to its final state $\psi_{m_1}$.

In this paper we address this problem and derive a theoretical expression for the inelastic magnetic scattering of neutrons by superconducting rings. The temperature of the ring is assumed to be small as compared with $T_c$ of the superconductor.

## 2. Hamiltonian of the system

The Hamiltonian of the system neutron – superconducting ring is:

$$H = \sum_m E_m a_m^+ a_m + \sum_{\mathbf{p},S} \varepsilon_{\mathbf{p}} c_{\mathbf{p}S}^+ c_{\mathbf{p}S} + \sum_{\mathbf{p},S,\mathbf{p}_1,S_1,m,m_1} V(\mathbf{p}S,\mathbf{p}_1 S_1; m, m_1) c_{\mathbf{p}_1 S_1}^+ c_{\mathbf{p}S} a_{m_1}^+ a_m , \qquad (1)$$

where $a_m^+(a_m)$ is the creation (destruction) operator of the $m$-state of the condensate in the ring; $\varepsilon_{\mathbf{p}}$, $\mathbf{p}$ and $S$ are, respectively, the initial energy, wave vector and spin of the neutron, $\mathbf{p}_1, S_1$ are the final wave vector and spin of the neutron. The third term on the right side of (1) is the matrix element of the interaction operator of neutron with the magnetic field of the superconducting ring, which has the form:

$$\hat{V} = 2\gamma \mu_N \hat{\mathbf{S}} \hat{\mathbf{B}}(\mathbf{r}_n) , \qquad (2)$$

where $\gamma = -1.93$; $\mu_N$ is the nuclear magneton; $\hat{\mathbf{S}}$ is the neutron spin operator; $\hat{\mathbf{B}}(\mathbf{r}_n)$ is the operator of the magnetic induction created by the ring current at the radius-vector of the neutron $\mathbf{r}_n$:

$$\hat{\mathbf{B}}(\mathbf{r}_n) = \frac{\mu_0}{4\pi} \int d\mathbf{r} \frac{\left[\hat{\mathbf{j}}(\mathbf{r}), \mathbf{r}_n - \mathbf{r}\right]}{|\mathbf{r}_n - \mathbf{r}|^3}, \qquad (3)$$

and $\hat{\mathbf{j}}$ is the current density operator in the superconducting ring.

Below we use the cylindrical coordinates ($z, \rho, \varphi$) with the $z$-axis perpendicular to the plane of the ring, and assume that the initial wave vector $\mathbf{p}$ of the neutron is directed along the $z$-axis. Then the scattering cross section is independent from the polar angle $\varphi$ due to the symmetry of the ring. In the first approximation of the Born scattering theory, the double differential cross section studied the inelastic process, is:

$$\frac{\partial^2 \sigma_{SS_1}}{\partial \varepsilon_{p_{1z}} \partial \varepsilon_{p_{1\rho}}} = 2^{-2} \pi^{-1} \frac{\Omega_n^2 m_n^2}{\hbar^4 \varepsilon_p^{1/2}} \sum_{m_1} \varepsilon_{p_{1z}}^{-1/2} | V_{\mathbf{p}\mathbf{p}_1}^{mm_1}(S, S_1)|^2 \delta(\varepsilon_{\mathbf{p}} - \varepsilon_{\mathbf{p}_1} + E_m - E_{m_1}) . \qquad (4)$$

Here $\varepsilon_{\mathbf{p}_1} = \dfrac{p_1^2}{2m_n} = \varepsilon_{p_{1z}} + \varepsilon_{p_{1\rho}}$ is the final energy of the neutron, $m_n$ is the neutron mass, $\Omega_n$ is the normalizing volume of the neutron wave functions.

Using functions of the plane waves, it is easy to calculate the matrix element of (2) over the initial state of the neutron $\psi_{\mathbf{p}}$ and its final state $\psi_{\mathbf{p}_1}$. As a result, we obtain:

$$\hat{V}_{\mathbf{pp}_1} = \int d\mathbf{r}_n \psi^*_{\mathbf{p}_1} \hat{V} \psi_{\mathbf{p}} = -2i\frac{\gamma\mu_0\mu_N}{\Omega_n q^2}\hat{\mathbf{S}}\Big[\mathbf{q}, \int d\mathbf{r}\,\hat{\mathbf{j}}(\mathbf{r})e^{i\mathbf{qr}}\Big], \qquad (5)$$

where $\mathbf{q} = \mathbf{p} - \mathbf{p}_1$ is the momentum transfer.

Using the well-known expression for the current density operator, its matrix element is written as:

$$\mathbf{j}_{mm_1} = \psi^*_{m_1}\hat{\mathbf{j}}\psi_m = \frac{ie\hbar}{m_C}\Big(\psi_m \nabla \psi^*_{m_1} - \psi^*_{m_1} \nabla \psi_m\Big) - \frac{4e^2}{m_C}\psi^*_{m_1}\hat{\mathbf{A}}\psi_m, \qquad (6)$$

where $m_C$ is the mass of the Cooper pair with its charge equal to $2e$, $\hat{\mathbf{A}}$ is the operator of the vector potential, which can be defined as:

$$\hat{\mathbf{A}}(\mathbf{r}) = \frac{\mu_0}{4\pi}\int d\mathbf{r}_1 \frac{\hat{\mathbf{j}}(\mathbf{r}_1)}{|\mathbf{r}-\mathbf{r}_1|}.$$

For the superconducting ring $\mathbf{j}_{mm_1} = j_{mm_1}(z,\rho)\mathbf{i}_\varphi$. Then the matrix elements of the operator (5) taken over the condensate wave functions of the initial $\psi_m$ and final $\psi_{m_1}$ states in the ring, are given by:

$$\hat{V}^{mm_1}_{\mathbf{pp}_1} = 2i\frac{\gamma\mu_0\mu_N}{\Omega_n q^2}\int d\mathbf{r}\, j_{mm_1}(z,\rho)e^{i\mathbf{qr}}\hat{\mathbf{S}}\big[\mathbf{i}_\varphi, \mathbf{q}\big]. \qquad (7)$$

Now we consider the matrix elements of (7) over the spin variables of neutron. For polarized neutrons we obtain:

$$<\alpha|\hat{\mathbf{S}}[\mathbf{i}_\varphi,\mathbf{q}]|\alpha> = -\frac{1}{2}q_\rho \cos(\varphi - \varphi_q),$$

$$<\beta|\hat{\mathbf{S}}[\mathbf{i}_\varphi,\mathbf{q}]|\beta> = \frac{1}{2}q_\rho \cos(\varphi - \varphi_q) \qquad (8)$$

for the scattering without the spin flip process, and

$$<\beta|\hat{\mathbf{S}}[\mathbf{i}_\varphi,\mathbf{q}]|\alpha> = \frac{1}{2}q_z e^{i\varphi}$$

$$<\alpha|\hat{\mathbf{S}}[\mathbf{i}_\varphi,\mathbf{q}]|\beta> = \frac{1}{2}q_z e^{-i\varphi} \qquad (9)$$

with the spin flip. Here $\varphi_q$ is the polar angle of the scattering vector $\mathbf{q}$, $|\alpha>$ and $|\beta>$ are the spin functions with $S_z$ equal to $+1/2$ and $-1/2$, respectively.

Substituting (8) and (9)-(11) in (4), the double differential cross section is reduced to the form:

$$\frac{\partial^2 \sigma}{\partial \varepsilon_{p_{1_z}} \partial \varepsilon_{p_{1_\rho}}} = 2^{-4} \pi^{-1} \gamma^2 \frac{e^2 \mu_0^2}{\hbar^2 \varepsilon_p^{1/2}} \sum_{m_1} \frac{j_{mm_1}^2(\mathbf{q})}{q^4 \varepsilon_{p_{1_z}}^{1/2}} F(q) \delta(\varepsilon_{\mathbf{p}} - \varepsilon_{\mathbf{p}_1} + E_m - E_{m_1}), \qquad (10)$$

where $F(q) = q_\rho^2$,

$$j_{mm_1}(\mathbf{q}) = \int d\mathbf{r} j_{mm_1}(z,\rho) \cos(\varphi - \varphi_q) e^{i\mathbf{qr}} \qquad (11)$$

for the scattering without the spin flip process, and $F(q) = q_z^2$,

$$j_{mm_1}(\mathbf{q}) = \int d\mathbf{r} j_{mm_1}(z,\rho) e^{i\mathbf{qr} \pm i\varphi} \qquad (12)$$

with the spin flip.

Formula (10) is a general expression for the inelastic scattering cross section of neutrons by superconducting rings. To analyze (10), the off-diagonal matrix elements of the operator of the superconducting current density in the ring are required.

Investigations of the current distributions corresponded to the diagonal matrix elements of the current density operator, in superconducting rings are known [5,6]. Typically, these results were obtained by numerical methods. However, these approaches do not allow clearly to analyze the dependence of the scattering cross section on the parameters of the superconductor, ring, and neutron energy. Below we consider a thin-film ring from the type-II superconductor, for which analytical expressions for the off-diagonal matrix elements of the superconducting current density can easily be obtained.

### 3. Off-diagonal matrix elements

Consider a rectangular cross-section ring from the type-II superconductor. The thickness of the ring obeys $d < \lambda$ (where $\lambda$ is the London magnetic penetration depth), its inner radius $a \gg \lambda$ and outer radius $b \gg a$. In the cylindrical coordinates ($z, \rho, \varphi$) the ring is in the region $-d/2 \leq z \leq d/2$. We assume that the magnetic field in the ring is weak as compared with $H_{c1}$ of the superconductor.

To find the magnetic induction in the ring, we use the London equation: $\Delta \mathbf{B} = \lambda^{-2} \mathbf{B}$. Since the ring thickness $d < \lambda$, the $z$-dependence of the magnetic induction can be neglected, and because of the circular symmetry, the magnetic field does not depend on the polar angle $\varphi$. The solution of the London equation is searched in the form: $\mathbf{B} = B_z(\rho)\mathbf{i}_z + B_\rho(\rho)\mathbf{i}_\rho$. As a result, for each component of the field we have the equation:

$$t^2 B_{tt}^{''} + t B_t^{'} - t^2 B = 0, \qquad (13)$$

where $t = \rho/\lambda$. In general, the solution of Eq. (13) is expressed in terms of the modified Bessel functions $I_o(\rho/\lambda)$ and $K_o(\rho/\lambda)$. The function $I_o(\rho/\lambda)$ increases exponentially with $\rho$. Since the outer radius of the ring $b \gg a$, the field at the outer surface of the ring can be neglected. Then, from Eq. (13) we obtain:

$$\mathbf{B}_z(\rho) = B_m K_0(\rho/\lambda)\mathbf{i}_z, \qquad (14)$$

where $B_m$ is a function of the quantum number $m$. The current density in the ring ($\mathbf{j} = \mu_0^{-1} rot \mathbf{B}$) is determined only by the $z-$component of the magnetic induction. Using (14), we have:

$$\mathbf{j}_m(\rho) = \frac{B_m}{\mu_0 \lambda} K_1(\rho/\lambda) \mathbf{i}_\varphi. \qquad (15)$$

Quantization of the magnetic flux trapped in the ring, allows to find $B_m$. Indeed, using (15), the superconducting current $J_m = \int \mathbf{j}_m d\mathbf{S}$, where the integration is taken over the cross section of the ring, is:

$$J_m = \frac{B_m d}{\mu_0} K_0(a/\lambda). \qquad (16)$$

Since the current (16) creates the magnetic induction flux $\Phi_m = \Phi_0 m$, we obtain:

$$B_m = \frac{\mu_0 \Phi_0}{dL K_0(a/\lambda)} m, \qquad (17)$$

where the quantum number $m$ is restricted by

$$m < m_{\max} = \frac{H_{c1} dL}{\Phi_0}.$$

The current distribution (15) can be regarded as the diagonal elements of the current density operator taken over the condensate wave function $\psi_m$: $\mathbf{j}_m = \psi_m^* \hat{\mathbf{j}} \psi_m$. The expression (6) allows to determine the off-diagonal matrix elements by the symmetric form:

$$\mathbf{j}_{mm_1} = \psi_{m_1}^* \hat{\mathbf{j}} \psi_m = \frac{1}{2}(\mathbf{j}_m + \mathbf{j}_{m_1}) \exp(i(\Phi_m - \Phi_{m_1})), \qquad (18)$$

where $\Phi_m = m\varphi$ is the phase of the condensate wave function $\psi_m$.

In the London theory, the energy of the ring which is composed of the magnetic energy and the kinetic energy of the current, is equal to:

$$E_m = E_0 m^2 = \frac{1}{2} L J_m^2, \qquad (19)$$

where the characteristic energy of the ring:

$$E_0 = \frac{\Phi_0^2}{2L}. \qquad (20)$$

**4. The cross section**

Substituting (15) and (17) in (18) and, then, using the definition of $j_{mm_1}(\mathbf{q})$ (11), for the neutron scattering without the spin flip process we derive:

$$j_{mm_1}(\mathbf{q}) = \zeta\, 2^2 \pi\, \frac{\Phi_0 (m+m_1)}{2dL} \frac{\sin(\frac{1}{2} q_z d)}{q_z q_\rho} T_{nsf}(q_\rho, a, \lambda, m-m_1), \qquad (21)$$

where $q_\rho = p_{1_\rho}$, $\zeta = e^{i(m-m_1)\varphi_q}(-1)^{(m-m_1-1)/2}$ is an insignificant factor, modulus of which is equal to 1, and

$$T_{nsf}(q_\rho, a, \lambda, m-m_1) = K_0^{-1}(a/\lambda) \int_{a/\lambda}^{\infty} t\, dt\, K_1(t) \frac{d}{dt} J_{m-m_1}(q_\rho \lambda t). \qquad (22)$$

Here $J_{m-m_1}(q_\rho \rho)$ is the Bessel function.

Using the same way and the definition (12), for the neutron scattering with the spin flip process we obtain:

$$j_{mm_1}(\mathbf{q}) = \zeta\, 2^2 \pi\, \frac{\lambda \Phi_0 (m+m_1)}{2dL} \frac{\sin(\frac{1}{2} q_z d)}{q_z} T_{sf}(q_\rho, a, \lambda, m-m_1), \qquad (23)$$

where $q_\rho = p_{1_\rho}$, $\zeta = e^{i(m-m_1 \pm 1)\varphi_q}(-1)^{(m-m_1 \pm 1)/2}$ and

$$T_{sf}(q_\rho, a, \lambda, m-m_1) = K_0^{-1}(a/\lambda) \int_{a/\lambda}^{\infty} t\, dt\, K_1(t) J_{m-m_1 \pm 1}(q_\rho \lambda t). \qquad (24)$$

Here the $\pm$ sings correspond with the signs in expressions (9).

According to (21) and (23), a characteristic change of the $z-$component of the scattering wave vector $q_z = p - p_{1z} \approx \frac{2\pi}{d}$. For the rings with the thickness $d \approx 10^3\, \overset{0}{A}$ that is half as large of typical values of the penetration depth ($\lambda = 2*10^3\, \overset{0}{A}$), even for cold neutrons with the energy $1\,\mu\text{эB}$ the wave vector $p \gg \frac{2\pi}{d}$. In fact, (21) and (23) predict the conservation of the $z-$component of the wave vector of the neutron. Therefore, in (21) and (23) we can use the replacement:

$$\int_{-d/2}^{d/2} dz\, e^{iq_z z} = 2\frac{\sin(\frac{1}{2} q_z d)}{q_z} \cong 2\pi \delta(q_z). \qquad (25)$$

Taking into account (10) and (25), we conclude that the neutron scattering with the spin flip process is suppressed, and the inelastic scattering ($m \neq m_1$) must be accompanied by the appearance of the transverse momentum of the neutron.

Using (19)-(22) and (25) and integrating (10) over the energy $\varepsilon_{p_{1\rho}}$, the total cross section for the neutron scattering without the spin flip process is given by:

$$\sigma = \left(\frac{\pi}{2}\gamma\right)^2 \frac{e^2 \mu_0^2}{dp m_n L} \sum_{m_1=1}^{m-1} \frac{m+m_1}{m-m_1} p_{1_\rho}^{-2} T_{nsf}(p_{1_\rho}, a, \lambda, m-m_1)^2, \qquad (26)$$

where the function $T_{nsf}$ (22) is determined at $q_\rho = p_{1_\rho}$ with the transverse momentum of the scattered neutron

$$p_{1_\rho}(m_1) = \frac{\sqrt{2m_n E_0}}{\hbar}\left(m^2 - m_1^2\right)^{1/2}. \qquad (27)$$

This momentum depends on the final number $m_1$ of the magnetic flux quanta trapped in the ring. The integral on the right side of (22) is easily calculated. Since the ring radius $a \gg \lambda$, the asymptote of the modified Bessel functions $K_{0,1}(x) \cong 1.2533 x^{-1/2} e^{-x}$ [7] can be used. Further, the minimum value of the argument of the Bessel functions $J_{m-m_1}$ is $p_{1_\rho} a$. For macroscopic rings $p_{1_\rho} a \gg 1$ and, consequently, the asymptotic expansion of the Bessel functions for large argument [7] can be applied. As a result, we obtain:

$$T_{nsf}^2 = \frac{2}{\pi} a p_{1_\rho} \left(1 + \lambda^2 p_{1_\rho}^2\right)^{-2} \left[\cos(a p_{1_\rho} - \varphi) - \lambda p_{1_\rho} \sin(a p_{1_\rho} - \varphi)\right]^2, \qquad (28)$$

where $\varphi = \frac{\pi}{2}(m - m_1 - \frac{1}{2})$. Substituting (28) in (26), finally, for the cross section we have:

$$\sigma = \frac{\pi}{2} \gamma^2 \frac{e^2 \mu_0^2 a}{d p m_n L} \sum_{m_1=1}^{m-1} \frac{m+m_1}{m-m_1} \frac{\left[\cos(a p_{1_\rho} - \varphi) - \lambda p_{1_\rho} \sin(a p_{1_\rho} - \varphi)\right]^2}{p_{1_\rho} \left(1 + \lambda^2 p_{1_\rho}^2\right)^2}. \qquad (29)$$

## 5. Numerical values of the cross sections

From (14) and (16) we find that the component of the self-inductance of the ring due to the magnetic flux through the region of the superconductor is:

$$L_r = 2\pi\mu_0 \frac{\lambda}{d} a. \qquad (30)$$

Using the above asymptote of the modified Bessel functions $K_{0,1}$ in (15) and (16), the component of self-inductance due to the magnetic flux through the hole of the ring in the plane $z_0$ ($|z_0| \leq d/2$) is given by:

$$L_h = \mu_0 \frac{a^2}{d} \int_0^\infty e^{-t} dt \int_0^\pi \cos(\varphi) d\varphi \ln\frac{\left[\alpha_-^2 + t^2 + 2\beta(\beta+t)(1-\cos(\varphi))\right]^{1/2} + \alpha_-}{\left[\alpha_+^2 + t^2 + 2\beta(\beta+t)(1-\cos(\varphi))\right]^{1/2} - \alpha_+}, \qquad (31)$$

where $\alpha_\pm = (d \pm 2z_0)/2\lambda$ and $\beta = a/\lambda$. The total self-inductance of the ring is $L = L_r + L_h$. Note that for the rings with $a \geq 1$ mm and $d < \lambda$ the self-inductance (31) depends weakly from $z_0$. Thus, for any values of $z_0$ the relative change of $L_h$ is less then $10^{-3}$.

Introducing the dimensionless coefficient $\xi = \frac{1}{4}\gamma^2 L_r / L$, the cross section (29) is reduced to the form:

$$\sigma = \sum_{m_1=1}^{m-1} \sigma_p(m_1),$$

where $\sigma_p(m_1)$ is the partial cross section for the final state of the neutron - superconducting ring system with the neutron energy equal to $E_0(m^2 - m_1^2) + \varepsilon_p$ and the number of fluxoids reduced by $(m - m_1)$,

$$\sigma_p(m_1) = \xi \sigma_0 \frac{m + m_1}{m - m_1} \frac{[\cos(ap_{1_\rho} - \varphi) - \lambda p_{1_\rho} \sin(ap_{1_\rho} - \varphi)]^2}{\lambda p_{1_\rho}(m_1)(1 + \lambda^2 p_{1_\rho}^2)^2}. \tag{32}$$

Here

$$\sigma_0(\varepsilon_p) = \frac{e^2 \mu_0}{p m_n} = \frac{0.8778}{\sqrt{\varepsilon_p}} barn,$$

and the neutron energy $\varepsilon_p$ is taken in eV.

According to (27) and (32), the partial cross section $\sigma_p \propto E_0^{-3/2}$. Therefore, the smaller the characteristic energy of the ring $E_0$, the greater the cross section. The ring radius dependences of the characteristic energy $E_0$ (20) and total self-inductance $L$ (30) - (31) are shown in Fig. 1. In the calculations we used the penetration depth $\lambda = 2*10^3 \, \overset{0}{A}$ that is typical for type-II superconductors such as $Nb_3Sn$, $V_3Ga$, for which $H_{c1} \approx 200$ G. The self-inductance increases and, consequently, the characteristic energy decreases with increasing the ring radius. The dimensionless coefficient $\xi$ in (32) changes weakly, as shown in the inset in Fig. 1. With decreasing the ring thickness these dependences become more favorable for increasing the scattering cross section. Therefore we conclude that to increase the cross section the thinner ring with larger inner radius should be used.

For macroscopic rings with the radii $a \geq 1$ mm the phase of the harmonic functions in (32) is very large, $ap_{1_\rho}(m_1) - \varphi >>> \pi$. With small changing the ring radius the partial cross section (32) oscillates. For example, the characteristic scale of these oscillations is $\Delta a = \pi / p_{1_\rho}(m_1 = m - 1) \cong 1.65 \, \overset{0}{A}$ at the characteristic energy $E_0 = 0.0731$ meV corresponding to the ring with $a = 5$ mm and $d = \lambda/3$, as shown in Fig. 1, and the number $m = 51$.

Of course, the real rings can be characterized by the mean radius and its standard deviation. Then, the scattering cross section should be averaged over the radius distribution. Assuming that roughness of the ring inner surface is much larger the oscillation scale, we have:

$$<\sigma_p(m_1)> = \frac{1}{2} \xi \sigma_0 \frac{m + m_1}{m - m_1} \frac{1}{\lambda^3 p_{1_\rho}^3(m_1)}. \tag{33}$$

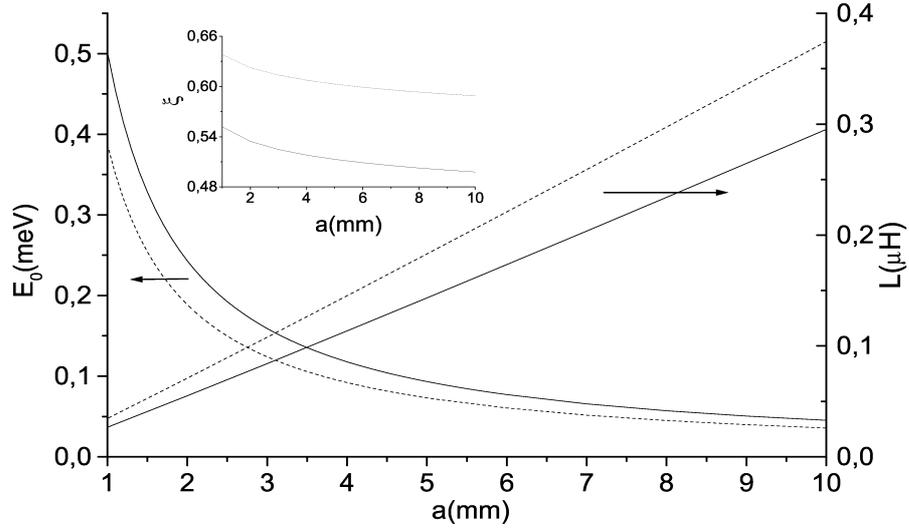

Fig. 1. Inner radius dependences of the characteristic energy $E_0$ and the self-inductance $L$ of the superconducting ring for the penetration depth $\lambda = 2*10^3 \; \overset{0}{A}$. The solid curves corresponds to the ring thickness $d = \lambda/2$, the dashed curves - $d = \lambda/3$. Inset: Inner radius dependence of the dimensionless coefficient $\xi$ in (32). The same notations are used.

In (33) there is the restriction on the number of fluxoids trapped in the ring, as discussed above. For the superconducting ring with $a = 5$ mm, $d = \lambda/3$ and $H_{c1} = 200$ G we obtain $m < m_{max} = 1.8*10^6$. The averaged cross section

$$<\sigma> = \sum_{m_1=1}^{m-1} <\sigma_p(m_1)> \qquad (34)$$

as a function of the initial number of fluxoids in the superconducting rings with three different radii is shown in Fig. 2. Obviously, the cross section increases with increasing the ring inner radius as is also with decreasing the number of fluxoids. A good approximation of the fluxoid number dependence of the cross section is $<\sigma> \propto m^{-1/2}$. Thus, to increase the inelastic scattering cross-section, the superconducting rings with relatively small numbers of magnetic flux quanta should be used.

The inset in Fig. 2 demonstrates the partial cross sections for the ring with $a = 10$ mm, $d = \lambda/3$ and $m = 51$. The main contribution to the total cross section is given by the channel $m_1 = m - 1$ that corresponds to the destruction of the one magnetic flux quantum in the superconducting ring and the neutron energy equal to $(2m-1)E_0 + \varepsilon_p$ after the scattering process. Also the channel $m_1 = m - 2$ for which the two magnetic flux quanta are destroyed in the superconducting ring and the neutron kinetic energy is increased by $4(m-1)E_0$, is significant.

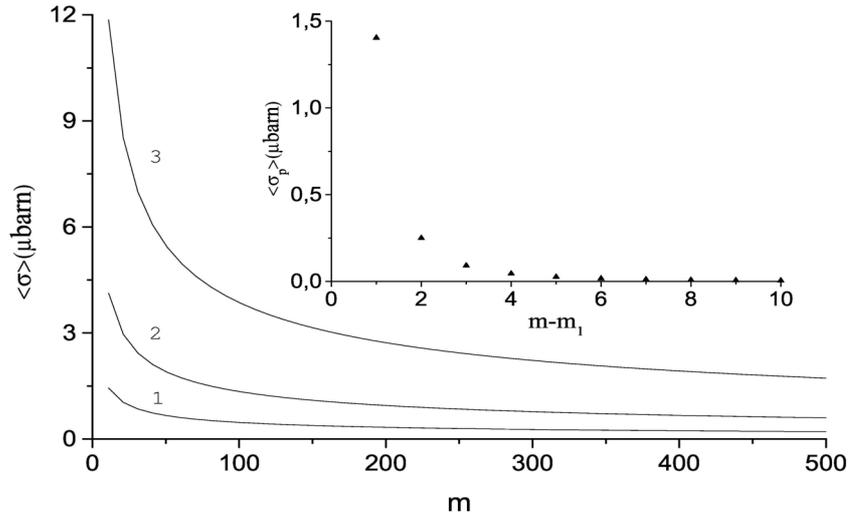

Fig. 2. The averaged cross section, (34), as a function of the initial number of fluxoids trapped in the superconducting rings, for three different inner radii: 1 - $a = 5$ mm, 2 - $a = 10$ mm and 3 - $a = 20$ mm. Inset: The partial cross-sections, (33), as a function of the scattering channels. The inner radius of the ring $a = 10$ mm and the initial number of fluxoids $m = 51$.

The angular distribution of the scattered neutrons does not depend on the polar angle $\varphi$, and is determined only by the discrete azimuthal angle $\theta(m_1) = artg\left(\sqrt{E_0(m^2 - m_1^2)/E_p}\right)$. The scattered neutrons move along the generating lines of conical surfaces. For the results shown in inset in Fig. 2, the two surfaces are deserved attention. The first cone defined by the angle $\theta = 88.50^0$, corresponds to the scattering channel $m_1 = m - 1$ for which the neutron energy is equal to $E_{p_1} = E_0(m^2 - m_1^2) + E_p = 3.607$ meV after the scattering, whereas the initial energy of the neutron is $E_p = 1 \mu$ eV. For the second conical surface defined by the angle $\theta = 88.93^0$ and corresponded to the channel $m_1 = m - 2$, the neutron energy is equal to $E_{p_1} = 7.141$ meV. That is, for the both channels the wave vector of the scattered neutrons is almost perpendicular to the initial wave vector.

## 6. Conclusions

In this paper a theory of inelastic magnetic scattering of neutrons by superconducting rings was developed taking account of the interaction of the neutron magnetic moment with magnetic field created by the superconducting current. It was shown that the differential cross section for the scattering process is defined by the off-diagonal matrix elements taken over the superconducting condensate wave functions in the ring, of the current density operator. For the thin-film rings from the type-II superconductors analytical expressions for the off-diagonal matrix elements of the superconducting current density were obtained. This is allowed to carry out calculations of the cross-section for cold neutrons as well as both energy and angle distributions of the scattered neutrons. We concluded that to increase the cross section, the thinner rings with larger inner radii and relatively small numbers of magnetic flux quanta trapped in the ring, should be used. These presented results we hope will stimulate experimental activity.

**References**

[1] Halpern O and Johnson M H  1939  *Phys. Rev.* **55**  898

[2] Izyumov Yu A 1997 *Phys. Usp.* **40** 521

[3] Lovesey S W  1977  *Dynamics of Solids and Liquids by Neutron Scattering*,  eds. S W Lovesey and T  Springer (Berlin: Springer-Verlag)

[4] Maleev S V 2002 *Phys. Usp.* **45** 569

[5] Pannetier M, Klaasen F C, Wijngaarden R J, Welling M. Heeck K, Huijbregtse J M, Dam B, and Griessen R  2001 *Phys. Rev.* B **64**  144505

[6] Brandt E H, Clem J R 2003 *Phys. Rev.* B **69**   184509

[7] *Handbook of mathematical functions* 1964 Appl. Math. Series 55, eds. M Abramowitz M and  I A Stegun  (Washington: National Bureau of Standards)